\documentclass[prd,onecolumn,showpacs,superscriptaddress,nofootinbib,floatfix,12pt,longbibliography]{revtex4-1} 
\usepackage[utf8]{inputenc}
\usepackage{color}
\usepackage{amsfonts,amsmath,amssymb}
\usepackage{graphicx}
\usepackage{multirow}
\usepackage{comment}
\usepackage{tkz-graph}
\usepackage{pifont}
\usepackage{slashed}
\usepackage{bm}
\usepackage{hyperref}
\hypersetup{colorlinks = true,
           allcolors = blue}
\usepackage{color}
\usepackage{epstopdf}
\usepackage{tabularx}

\def\x4c/{$X_{cc\bar c \bar c }$}
\def\xyy/{$\psi(4360)$}
\def\ppsi/{$\psi(4415)$}
\def\doned/{$D_1\bar{D}$}
\def\ddstar/{$D^* \bar{D}^*$}
\def\dtwodstar/{$D_2^*\bar{D}^*$} 
\def\dd/{$D \bar{D}$}
\def\*{^{(*)}}
\def\>{\big>}
\def\<{\big<}
\def\|{\big\vert}

\def\S{\mathbf S}
\def\s{\boldsymbol{\sigma}}
\def\three{\mathbf 3}
\def\six{\mathbf 6}
\def\k{\kappa}
\def\c{\bar c}
\def\b{\bar b}
\def\u{\bar u}
\def\d{\bar d}

\def\q{\bar q}
\def\Q{\bar Q}
\def\f{\varphi}
\def\l{\bm{\lambda}}
\def\jp{J/\psi}
\def\M{\overline M}
\def\H{\overline H}

\def\be{\begin{equation}}
\def\ee{\end{equation}}
\def\ba{\begin{eqnarray}}
\def\ea{\end{eqnarray}}

\renewcommand{\arraystretch}{1.25}

\begin{document}

\author{Muhammad Naeem Anwar}\email{m.n.anwar@swansea.ac.uk}
\author{Timothy J. Burns}\email{t.burns@swansea.ac.uk}
\affiliation{Department of Physics, Swansea University, Singleton Park, Swansea, SA2 8PP, UK.}

\title{Tetraquark mass relations in quark and diquark models}

\begin{abstract}
We present new linear relations among the masses of S-wave tetraquarks with either one flavour ($QQ \bar Q \bar Q$) or two ($QQ\bar q \bar q$). Because the relations are sensitive to the hidden-colour, spin, and spatial degrees of freedom, comparison to experimental data can help to reveal the internal structure of tetraquarks, and discriminate among different theoretical models. Depending on the model, the relations are either exact, or valid in perturbation theory, and a thorough comparison with existing literature confirms their validity at the MeV level. Additionally, we explore the connections among tetraquark models, and show how those with effective (quark or diquark) masses are related to dynamical potential models. We also show how the spectrum of diquark models is effectively a limiting case of (more general) quark models, and in particular, that the diquark concept is most relevant in the particular combination $QQ\bar q \bar q$, where $Q$ is much heavier than $\bar q$.
\end{abstract}

\date{\today}
\maketitle

\vspace{2cm}

\section{Introduction}

The ``November Revolution" provoked by the discovery of $J/\psi$~\cite{SLAC-SP-017:1974ind,E598:1974sol} is now sometimes known as the ``first'' charm revolution, owing to a more recent sequence of discoveries which are also deserving of revolutionary status.  A characteristic feature of the ``second'' charm revolution, which started at BaBar and Belle, and is still ongoing at BESIII and the LHC experiments at CERN, is the discovery of states which cannot apparently be understood as ordinary $q\q$ mesons or $qqq$ baryons; the experimental situation is reviewed in Refs.~\cite{Lebed:2016hpi,Brambilla:2019esw,Chen:2022asf}.
The new class of hadrons, which includes states in the charm and bottom quark sector, poses a significant challenge to our understanding of the strong interaction.
Future experiments, such as Belle II~\cite{Belle-II:2018jsg}
and PANDA~\cite{PANDA:2009yku}, are designed to further explore these hadrons.

The initial flood of new states became known collectively as the ``\textit{XYZ}'' states, a name strongly suggestive of their mysterious characteristics. Some of the states are by now so well-established that their nomenclature reflects the standard conventions of the Particle Data Group (PDG)~\cite{ParticleDataGroup:2022pth} -- so, for example, the state $X(3872)$~\cite{Belle:2003nnu} which launched the second charm revolution is now known as $\chi_{c1}(3872)$. Even so, the underlying nature of many of these states is not well understood, and there is considerable ongoing theoretical debate~\cite{Burns:2015dwa,Burns:2016gvy,Lebed:2016hpi,Ali:2017jda,Lu:2017yhl,Brambilla:2019esw,Liu:2019zoy,Hanhart:2019isz,Burns:2020epm,Burns:2020xne,Burns:2021jlu,Dong:2021bvy,Wu:2022bgq,Burns:2022uha}.

A key dividing line in these discussions is between molecular models and ``compact'' multiquark models, which have characteristically different degrees of freedom. In molecular models, the constituents are hadrons, whose interactions can be modelled, for example, in terms of pion exchange, or as effective field theory contact terms which are fit to data~\cite{Tornqvist:1991ks,Tornqvist:1993ng,Swanson:2003tb,Swanson:2004pp,Thomas:2008ja,FernandezCarames:2009zz,Guo:2017jvc,Burns:2019iih,Burns:2020xne,Anwar:2021dmg,Burns:2022uiv}. Such approaches are essentially an extension into the heavy quark sector of ideas which are widely applied in nuclear physics.

The focus of this paper is instead on compact multiquarks, which, by comparison to molecular models, are more ``exotic'', in the sense that there is no effective description in terms of interacting hadrons. Instead, taking a $QQ\q\q$ tetraquark as an example, the relevant degrees of freedom are typically assumed to be four interacting quarks, or alternatively,  effective $QQ$ and $\q\q$ diquarks. (Here $Q$ and $q$ are not necessarily heavy and light quarks, but rather, any distinct quark flavours.) One of our main motivations in this paper is to distinguish between these two physical pictures, which we refer to as quark models and diquark models, respectively.
(We give citations to the relevant literature in the main body of the paper, where the different models are described in more detail.)

Models for compact multiquarks have parameters which are typically not well-constrained, as they are usually fixed by comparison to the spectrum of conventional mesons and baryons, which introduces a systematic uncertainty which is difficult to quantify. Absolute predictions for the masses of states are therefore not very reliable, and moreover, they cannot be used to distinguish between quark and diquark models, whose predictions are similar within (large) uncertainties. By contrast, predictions for relations among masses (or mass splittings) are more general and, in some cases, are completely independent of parameters. Such relations, which are the main focus of this paper, obviously have more predictive power, and allow for more direct tests of model assumptions.

For conventional hadrons, the Gell-Mann–Okubo formula~\cite{Gell-Mann:1961omu,Gell-Mann:1962yej, Okubo:1961jc,Okubo:1962zzc} is a prototypical example of an empirically successful relation among hadron masses. Additional relations among the masses of conventional mesons and baryons have also been discovered and compared favourably to experimental data, for example in Refs.~\cite{Lipkin:1978ie,Cohen:1980su,Lipkin:1986dx,Lipkin:1989zq,Lipkin:1989pp,Karliner:2006fr,Karliner:2008sv}. In this paper we uncover similar relations among the masses of tetraquark states, and since they are based on similar symmetry arguments, we expect them to be equally reliable.

An analogy with ordinary $Q\Q$ mesons is instructive. In that case, absolute mass predictions (in quark potential models) have considerable uncertainty, but a linear relation among the masses in the $P$-wave multiplet is very reliable and is satisfied in experiments to less than an MeV \cite{Voloshin:2007dx,Burns:2011fu,Burns:2011jv,Burns:2014zfa,Burns:2014qya},
and also served as a benchmark for exotic structures in that mass region~\cite{Lebed:2017yme,Anwar:2018yqm}. The relations we find in this paper are conceptually very similar. Note that in this paper we are concentrating on relations \textit{among} the masses of tetraquark states, as distinct from relations between their masses and those of conventional mesons, which have also been discussed in the literature~\cite{Lipkin:1986dw,Eichten:2017ffp,Anwar:2017toa}, but remain to be confirmed experimentally. 

Our main results in this paper are for tetraquarks with either two flavours in the combination $QQ\q\q$, or just one flavour $QQ\Q\Q$.
(For specific examples, $Q$ can be regarded as a heavy quark flavour; however, $q$ is not necessarily a light quark, but rather, a distinct quark flavour from $Q$.)
Our interest in these particular combinations is partly due to recent experimental discoveries and lattice calculations. The one-flavour case has been a particularly hot topic recently, owing to a sequence of experimental observations showing apparent candidates for $cc\c\c$ states in $\jp\jp$ decays \cite{LHCb:2020bwg,CMS:2023owd,ATLAS:2023bft}; many of the results of this paper can be usefully applied to the phenomenology of these states~\cite{Anwar:2023fbp,*Anwar2023}.
Similarly, there is a growing body of evidence from experiment, lattice QCD and models,  for the likely binding of $bb\q\q'$ and $cc\q\q'$ tetraquarks, where $\q$ and $\q'$ are  light flavours \cite{Lipkin:1986dw,Zouzou:1986qh,Silvestre-Brac:1993zem,Silvestre-Brac:1993wyf,Francis:2016hui,Bicudo:2016ooe,Czarnecki:2017vco,Karliner:2017qjm,Eichten:2017ffp,Mehen:2017nrh,Francis:2018jyb,Junnarkar:2018twb,Leskovec:2019ioa,Mohanta:2020eed,Lu:2020rog,LHCb:2021vvq,LHCb:2021auc,Hudspith:2023loy,Aoki:2023nzp}. Note however that our results would only apply where  $\q$ and $\q'$ are identical or, from isospin symmetry, are an isovector $\u\d$ combination. 

The possible mixing between compact tetraquark and molecular state (such as explored in~\cite{Niu:2022jqp,Peng:2023lfw}) is not considered here. A possible caveat is that the potential for such a mixing introduces some additional parameter(s) and the derived mass relations will be subjected to some level of model dependence.

We begin (Section~\ref{sec:colour}) with some remarks on the main distinguishing feature of quark and diquark models, namely the assumed colour wavefunctions. We then
discuss quark models (Section~\ref{sec:quark}), showing that the chromomagnetic quark model (with effective quark masses) can be obtained in a symmetry limit from the quark potential model. We then do a similar exercise for diquark models (Section~\ref{sec:diquark}), and show how these are related to quark models in a truncated basis of colour. Specialising to tetraquarks with one or two flavours (Section~\ref{sec:massformulae}), we derive formulae for the masses of states in both quark and diquark models, and show how the diquark model emerges as a limiting case of quark models. Using the mass formulae, we identify linear relations among tetraquark masses (Section~\ref{sec:massrelations}), and show how these can discriminate among models. Finally (Section~\ref{sec:conclusion}) we summarise our results and suggest how they may  be used to inform comparisons with emerging experimental data.

\section{Colour, spin and flavour}
\label{sec:colour}

The key distinction between quark and diquark models is the treatment of colour. A pair of quarks can be coupled to colour $\bar\three$ or $\six$, while a pair of antiquarks can be coupled to $\three$ or $\bar\six$. To form an overall colour singlet, the possible combinations are then $\bar\three\otimes\three$ or $\six\otimes\bar\six$. A basic assumption of quark models is that both possibilities should be considered, and in general, a quark model state can be an admixture of the two. In diquark models, by construction, only the $\bar\three\otimes\three$  configuration is included~\cite{Lichtenberg:1996fi,Jaffe:2004ph}. 

As well as the treatment of colour, models are also distinguished according to whether the constituents (quarks or diquarks) have effective masses, or instead are dynamical objects whose contribution to the tetraquark mass is obtained from the Schr\"{o}dinger equation with some confining potential. We will consider both of these approaches, and the relation between them.

In this paper we concentrate on tetraquarks with either two flavours (in the combination $QQ\q\q$), or one ($QQ\Q\Q$). Both systems are subject to the same constraints, from the Pauli principle, on the allowed spin and colour configurations. With reference to the $QQ\q\q$ system, an S-wave $QQ$ pair can have (colour, spin) quantum numbers ($\bar\three$,1) or ($\six$,0), while an S-wave $\q\q$ pair can be  ($\three$,1) or ($\bar\six$,0). Forming an overall colour singlet, and combining the spins in S-wave to angular momentum $J$, the allowed combinations (and their $J^{P(C)}$ quantum numbers) are
\begin{align}
\|\f_2\>&=\|\{(QQ)_{\bar\three}^1(\q\q)_\three^1\}^2\> \quad [2^{+(+)}],\label{eq:basis2}\\
\|\f_1\>&=\|\{(QQ)_{\bar\three}^1(\q\q)_\three^1\}^1\> \quad [1^{+(-)}],\label{eq:basis1}\\
\|\f_0\>&=\|\{(QQ)_{\bar\three}^1(\q\q)_\three^1\}^0\> \quad [0^{+(+)}],\label{eq:basis0}\\
\|\f_0'\>&=|\{(QQ)_{\six}^0(\q\q)_{\bar\six}^0\}^0\> \quad [0^{+(+)}],\label{eq:basis0p}
\end{align}
where on the right-hand side, the subscripts are colour, and superscripts are spin. The charge conjugation quantum number $C$ is relevant only for the one-flavour case, corresponding to $Q=q$. 

When counting the number of distinct quark flavours, we can treat $u$ and $d$ quarks as identical if they come in the isovector (symmetric) combination, since they are subject to the same constraints from the Pauli principle outlined above. So, for example, results we obtain for $(I,I_3)=(1,\pm 1)$ states $QQ\d\d$ and $QQ\u\u$ apply equally to the $(I,I_3)=(1,0)$ partner $QQ\u\d$, but would not apply to an $(I,I_3)=(0,0)$ partner.

Diquark models are characterised by the inclusion of only colour triplet combinations, meaning the spectrum has three states ($\f_2$, $\f_1$ and $\f_0$). Quark models, by contrast, include both the colour triplet and colour sextet combinations, so there are four states, namely $\f_2$, $\f_1$, and two scalars, which are admixtures of $\f_0$ and $\f_0'$. Obviously an experimental determination of the number of scalar states can distinguish diquark models (one state) from quark models (two).

\section{Quark models}
\label{sec:quark}

In the chromomagnetic quark model, also known as the colour-magnetic interaction (CMI) model, the quark constituents have effective (rather than dynamical) masses, and the splitting among the S-wave states is  induced by chromomagnetic interactions (one-gluon exchange) \cite{DeRujula:1975qlm,Jaffe:1976ig,Jaffe:1976ih,Mulders:1979ea,Aerts:1979hn,Hogaasen:2005jv,Buccella:2006fn,Cui:2006mp,Stancu:2009ka,Hogaasen:2013nca,Karliner:2016zzc,Wu:2016vtq,Stancu:2016sfd,Cheng:2020wxa,Weng:2020jao}. The model has been widely applied to exotic hadron spectroscopy, as reviewed in Ref.~\cite{Liu:2019zoy}. The Hamiltonian for S-wave states is
\be
H=\M
- \sum_{i<j} C_{ij}~\l_i\cdot\l_j~\s_i \cdot\s_j,\label{eq:h:chromo}
\ee
where the centre of mass
\begin{align}
    \M=\sum_i m_i
\end{align}
is the sum of quark masses, $\l_i$ and $\s_i$ are the $SU(3)$ colour and $SU(2)$ spin (Pauli) matrices of quark $i$, and $C_{ij}$ are (positive) parameters which depend on quark flavours. The eigenstates of $H$ are, in general, admixtures of $\bar\three\otimes\three$ and $\six\otimes\bar\six$ colour configurations, with mixing induced by the $\l_i\cdot\l_j$ term. 

The parameters $\M$ and $C_{ij}$ are typically fixed by applying the same Hamiltonian  to the spectrum of conventional mesons and/or baryons, and fitting. An explicit assumption is that the same coefficients $C_{ij}$ control the interactions between any pair of flavours $i$ and $j$, either as a quark-quark ($q_iq_j$) or quark-antiquark ($q_i\q_j$) pair\footnote{$q_{i,j}$ stands for any quark flavour either heavy or light throughout.}, and regardless of whether these pairs are in a tetraquark, or in a conventional meson or baryon. In some cases, it is further assumed that the coefficients $C_{ij}$ scale inversely with quark masses,
\begin{align}
    C_{ij}=\frac{c}{m_im_j},\label{eq:cij1}
\end{align}
for some constant $c$ (which will be identified later in this section).

Quark potential models \cite{Ader:1981db,Weinstein:1982gc,Weinstein:1983gd,Godfrey:1985xj,Zouzou:1986qh,Weinstein:1990gu,Silvestre-Brac:1993zem,Silvestre-Brac:1993wyf,Brink:1994ic,Brink:1998as,Lloyd:2003yc,Vijande:2003ki,Barnea:2006sd,Vijande:2006dq,Vijande:2007rf,Vijande:2007fc,Vijande:2009zs,Vijande:2009kj,Vijande:2009ac,Lu:2016mbb,Anwar:2017toa,Richard:2017vry,Richard:2018yrm,Wang:2019rdo,Hernandez:2019eox,Liu:2019zuc,Wang:2022yes,Zhang:2022qtp} are a widely-used (and somewhat more rigorous) alternative approach, in which the quark constituents are dynamical, rather than having effective masses. A typical quark model Hamiltonian
\begin{align}
    H=T+V+U,
\end{align}
has a potential with chromoelectric ($V$) and chromomagnetic ($U$) contributions
\begin{align}
    V&=-\sum_{i<j}\l_i\cdot\l_j ~v(r_{ij}),\\
    U&=-\sum_{i<j}\l_i\cdot\l_j ~\s_i \cdot\s_j ~u(r_{ij}),
\end{align}
whose radial parts are typically (but not necessarily) of the form
\begin{align}
    v(r_{ij})&=\frac{3}{16}\left(b\,r_{ij}
-\frac{4}{3}\frac{\alpha_s}{r_{ij}}+c_0\right),\\
    u(r_{ij})&= \frac{\pi}{6}\, \frac{ \alpha_s}{m_i m_j} \, \delta^3 (r_{ij}) \, ,\label{eq:u}
\end{align}
where $b$ and $\alpha_s$ are the strengths of the confining force (string tension) and color-Coulomb potential, respectively, and $c_0$ is mass renormalization. Numerical values of these parameters can be extracted from the hadron spectrum~\cite{Godfrey:1985xj,Lu:2016mbb,Lu:2020rog,Zhang:2022qtp}.

Comparing the two models, it is clear that $U$ in the potential model is closely related to the interaction term in the chromomagnetic model \eqref{eq:h:chromo}. To understand the relationship between the models, we treat $U$ as a perturbation and consider the Hamiltonian
\begin{align}
    \overline H=T+V \, ,
\end{align}
whose eigenstates are $\f_0$, $\f_1$, $\f_2$ and $\f_0'$  introduced in equations \eqref{eq:basis2}-\eqref{eq:basis0p}. (There is no term in $\overline H$ which mixes $\f_0$ and $\f_0'$, due to the orthogonality of the spin wavefunctions.) Because $\overline H$ depends on colour but not spin, there is degeneracy among the states $\f_0$, $\f_1$, and $\f_2$, but not between these and $\f_0'$, 
\begin{align}
    \<\f_J\|\overline H\|\f_J\>\ne
    \<\f_0'\|\overline H\|\f_0'\> \, .\label{eq:nondegenerate}
\end{align}
We will point out that in order to make the connection with the chromomagnetic model, an extra symmetry constraint is required, which restores this degeneracy. 

The matrix elements of $V$, using the colour matrix elements in Ref.~\cite{Vijande:2009ac}, are
\begin{align}
\<\f_J\|V\|\f_J\>&=\frac{8}{3}\<v(r_{12})+v(r_{34})\>+
\frac{4}{3}\<v(r_{13})+v(r_{14})+v(r_{23})+v(r_{24})\> \, ,\\
\<\f_0'\|V\|\f_0'\>&=-\frac{4}{3}\<v(r_{12})+v(r_{34})\>+
\frac{10}{3}\<v(r_{13})+v(r_{14})+v(r_{23})+v(r_{24})\>
\end{align}
where an integral over all spatial degrees of freedom is implied. From the symmetries in $T$ and $V$, the ground state wavefunctions are symmetric under the interchange of quarks~($1\leftrightarrow 2$), antiquarks ($3\leftrightarrow 4$), or both $(12\leftrightarrow 34)$, so the spatial integral can be reduced to two independent terms
\begin{align}
\<\f_J\|V\|\f_J\>&=\frac{16}{3}\<v(r_{12})\>+\frac{16}{3}\<v(r_{13})\> \, ,\\
\<\f_0'\|V\|\f_0'\>&=-\frac{8}{3}\<v(r_{12})\>+\frac{40}{3}\<v(r_{13})\>\, .
\end{align}
Note however that the wavefunction does not have an additional symmetry under the interchange of a quark and antiquark (such as $2\leftrightarrow 3$), so in general, no further simplification is possible. This applies not only to states with two flavours ($QQ\q\q$), but also states with one flavour ($QQ\Q\Q$): the Hamiltonian does not impose a symmetry under $Q\leftrightarrow\q$ or $Q\leftrightarrow\Q$, so the wavefunction does not have that symmetry. It turns out, however, that this additional symmetry is often imposed on the wavefunction as an artefact of the calculation. In particular this is often the case when the Gaussian Expansion Method is applied to tetraquarks with one flavour ($QQ\Q\Q$), as in for example Refs.~\cite{Liu:2019zuc,Zhang:2022qtp}. If we impose this extra symmetry (under $2\leftrightarrow 3$), the spatial integral reduces further, and the potentials for all states are identical
\begin{align}
\<\f_J\|V\|\f_J\>=
\<\f_0'\|V\|\f_0'\>=\frac{32}{3}\<v(r_{12})\>\, ,
\end{align}
which further implies, as distinct from the general case \eqref{eq:nondegenerate}, that the eigenstates of $\overline H$ are degenerate. Identifying $\M$ in equation~\eqref{eq:h:chromo} as the corresponding eigenvalue,
\begin{align}
\<\f_J\|\overline H\|\f_J\>=
    \<\f_0'\|\overline H\|\f_0'\>\equiv \M \, ,
\end{align}
we see that a perturbative treatment of the full quark model Hamiltonian
\begin{align}
    H=\overline H+U \, ,\label{eq:pert}
\end{align}
is equivalent to the chromomagnetic model \eqref{eq:h:chromo}, where the coefficients $C_{ij}$ are obtained from $u(r_{ij})$ by integrating over the spatial wavefunctions of the eigenstates of $\overline H$,
\begin{align}
    C_{ij}=\<u(r_{ij})\> \, .\label{eq:cij2}
\end{align}

Note that with this interpretation, the centre of mass term $\M$ is no longer just the sum of quark masses, but also absorbs the dynamical contributions from the potential model,  namely the kinetic energy and the confining term. Also, the coefficients $C_{ij}$ depend not only on quark masses, as in equation~\eqref{eq:cij1}, but also depend on the spatial wavefunction of the quarks.

In the symmetry limit we are working in, $\<r_{ij}\>$ is independent of $i$ and $j$, hence so is $C_{ij}$. This validates the assumption, in the chromomagnetic model, that the same $C_{ij}$ can be used for any pair of flavours $i$ and $j$ in a tetraquark, both quark-quark ($q_iq_j$) and quark-antiquark ($q_i\q_j$) pairs. However it does not establish that one can use the same $C_{ij}$ in tetraquarks as in conventional mesons and baryons: this remains an assumption of the model, which could however be tested by evaluating equation \eqref{eq:cij2} and the corresponding expression for mesons and baryons.

With the specific form of $u(r_{ij})$ in equation \eqref{eq:u}, we reproduce the inverse dependence of $C_{ij}$ on quark masses as in equation~\eqref{eq:cij1}, with
\begin{align}
    c=\frac{\pi}{6}\,\alpha_s \<\delta^3 (r_{ij})\>\label{eq:cij3} \, .
\end{align}
Note that, in the symmetry limit we are working in, $c$ is indeed constant, in the sense of being independent of the flavours $i$ and $j$ within the tetraquark.

\section{Diquark models}
\label{sec:diquark}

A widespread implementation of the diquark model \cite{Maiani:2004vq,Maiani:2005pe,Drenska:2008gr,Drenska:2009cd,Ali:2009pi,Ali:2011vy,Ali:2019npk} has a Hamiltonian which is very similar to that of the chromomagnetic model,
\begin{align}
	H=\M +2\sum_{i<j}\k_{ij}~\S_i\cdot\S_j \, ,\label{eq:h:diq}
\end{align}
where $\M$ is the sum of quark or diquark effective masses, $\S_i=\s_i/2$ is the spin of quark $i$, and $\k_{ij}$ are (positive) parameters which depend on quark flavours and which, unlike the parameters $C_{ij}$ of the chromomagnetic model, are not assumed to be the same for quark-quark ($q_iq_j$) and quark-antiquark ($q_i\q_j$) combinations. Otherwise, the only distinction between the diquark model and the chromomagnetic model is the use of a truncated colour basis $\bar\three\otimes\three$. If we evaluate the chromomagnetic Hamiltonian \eqref{eq:h:chromo} in the same basis, the two models are equivalent provided their couplings are related
\begin{align}
    \k_{ij}=-2\<\l_i\cdot\l_j\>C_{ij}
\end{align}
namely 
\begin{equation}
    \k_{ij}=\left\{\begin{matrix}\frac{16}{3}C_{ij},&\textrm{ for }q_iq_j ,\\
    \frac{8}{3}C_{ij},&\textrm{ for }q_i\q_j,\end{matrix}\right .
\end{equation}
where we have used the colour matrix elements in Ref.~\cite{Vijande:2009ac}. In this sense, the diquark model is identical to the chromomagnetic quark model, but evaluated in a truncated colour basis. We will later use this property to extract the spectrum of the diquark model as a limiting case of the chromomagnetic quark model.

Referring to the Hamiltonian \eqref{eq:h:diq} as a diquark model is somewhat counterintuitive, since the spin degrees of freedom are actually quarks (not diquarks). The distinction turns out not to be important for the particular flavour combinations of tetraquarks which are the focus of this paper. For $QQ\q\q$ states there are three independent couplings
\begin{align}
    \k_{QQ}&\equiv\k_{12} \, ,\\
	\k_{qq}&\equiv\k_{34} \, ,\\
	\k_{Q\q}&\equiv\k_{13}=\k_{14}=\k_{23}=\k_{24} \, ,\label{eq:kQq}
\end{align}
with the obvious simplification to two couplings ($\k_{QQ}$ and $\k_{Q\Q}$) in the special case $QQ\Q\Q$. We consider the (more general) $QQ\q\q$ case in detail. With the couplings above, the Hamiltonian reduces to
\begin{align}
    H=\M +\frac{1}{2}\left(\k_{QQ}+\k_{qq}\right)+2\k_{Q\q}~\S_{12}\cdot\S_{34} \, ,\label{eq:diq:eff}
\end{align}
where here we have evaluated
\begin{align}
    \<\S_{1}\cdot\S_2\>=\<\S_{3}\cdot\S_4\>=\frac{1}{4} \, ,
\end{align}
as appropriate to spin-1 diquarks. The key feature is that the spin-dependence of the Hamiltonian is now expressed in terms of effective diquark spin operators
\begin{align}
	\S_{12}&=\S_1+\S_2\label{eq:effd1} \, ,\\
	\S_{34}&=\S_3+\S_4\label{eq:effd2}\, ,
\end{align}
corresponding to the total spin of the $QQ$ and $\q\q$ diquarks, respectively. In this sense, the Hamiltonian defined at quark level can be naturally interpreted in terms of diquark degrees of freedom. But this is a peculiarity of the flavour combination $QQ\q\q$ (or $QQ\Q\Q$), which leads to equation~\eqref{eq:diq:eff}. For other combinations of flavours, such as $QQ\bar Q\q$ and $Qq\bar Q\q$, the same does not apply, and in general an effective diquark description does not emerge in the same way; this is because as well as effective diquark operators like \eqref{eq:effd1} and \eqref{eq:effd2}, there are other operators $\S_1-\S_2$ and/or $\S_3-\S_4$ which mix ``diquarks'' with different spin.

Diquark potential models are more explicit in treating diquarks as effective degrees of freedom. Here the mass spectrum of tetraquark states comes from the Schr\"{o}dinger equation, in which diquarks are massive (colour-triplet) objects interacting through a confining  potential~\cite{Ebert:2005nc,Ebert:2007rn,Ebert:2008kb,Ebert:2008se,Ebert:2010af,Faustov:2020qfm,Anwar:2018sol,Bedolla:2019zwg,Anwar:2017toa,Debastiani:2017msn,Anwar:2018sol,Lundhammar:2020xvw}. The distinctions with the previous model are that the mass spectrum comes from the Schr\"{o}dinger equation (rather than effective masses fit to data), and that the spin dependence is expressed from the outset in terms of diquark (not quark) spin operators.

To clarify the relation between the different diquark models, we follow a similar procedure to our previous discussion of quark models. In the Hamiltonian,
\begin{align}
    H=T+V+U \, ,\label{eq:diq:full}
\end{align}
we isolate the spin-independent kinetic ($T$) and confining ($V$) terms, and a spin-dependent term ($U$), which in all models is expressed (for S-wave states) in terms of 
diquark spin operators, 
\begin{align}
    U=u(r)~\S_{12}\cdot\S_{34}\, ,
\end{align}
where here $r$ is the radial component of the vector joining the diquark and antidiquark.
There is considerable variation among the different approaches to diquark potential models, for example, in how the effective diquark mass is obtained, the assumed form of the confining potential $V$, and the precise form of the radial component $u(r)$ of the spin-spin term. However these differences are immaterial to the discussion.

As in the quark model case, we treat $U$ as a perturbation. If we identify the eigenvalues of the spin-independent part
\begin{align}
    \H=T+V
\end{align}
with the spin-independent term in equation~\eqref{eq:diq:eff}, 
\begin{align}
\< \H\>\equiv \M+\frac{1}{2}\left(\k_{QQ}+\k_{qq}\right) ,
\end{align}
we notice that a perturbative treatment of the full Hamiltonian \eqref{eq:diq:full} is equivalent to the previous diquark model, with the couplings defined as integrals over the eigenstates of $\H$,
\begin{align}
    \k_{Q\q}\equiv\frac{1}{2}\<u(r)\> \,.
\end{align}

\section{Mass formulae}
\label{sec:massformulae} 

At this stage we have established the underlying connections among four different classes of models, distinguished according to the colour structure (quarks versus diquarks), and the mass spectrum (effective masses versus dynamical masses from the Schr\"{o}dinger equation). Among the models with effective masses, we showed that the quark model \eqref{eq:h:chromo} and diquark model \eqref{eq:h:diq} are equivalent, except that the latter is evaluated in a truncated colour basis. We also showed that each of these models can be understood by applying perturbation theory to a corresponding (quark or diquark) potential model, which gives a dynamical interpretation for the effective masses, and implies that the couplings are sensitive to the spatial wavefunctions. 

Having established these connections among models, we will now obtain some general results which apply to all four classes of model. As a framework for our calculation,  we will use the most general Hamiltonian \eqref{eq:h:chromo}. By evaluating its spectrum in the full and truncated colour basis, respectively, we get results which correspond to quark and diquark models. 

When applying the Hamiltonian \eqref{eq:h:chromo} to $QQ\q\q$ states, there are three possible couplings,
\begin{align}
	C_{QQ}&=C_{12}\label{eq:C1}\, ,\\
	C_{qq}&=C_{34}\label{eq:C2}\, ,\\
	C_{Q\q}&=C_{13}=C_{14}=C_{23}=C_{24}\, ,\label{eq:C3}
\end{align}
whereas for $QQ\Q\Q$ states there are only two,
\begin{align}
	C_{QQ}&=C_{12}=C_{34} \, ,\\
	C_{Q\Q}&=C_{13}=C_{14}=C_{23}=C_{24}\, .
\end{align}
We will discuss the more general $QQ\q\q$ case, noting that $QQ\Q\Q$ is then a special case with $Q=q$.

We need the matrix elements of $H$ with respect to the basis states \eqref{eq:basis2}-\eqref{eq:basis0p}. Using, for example, the colour and spin matrix elements of Ref.~\cite{Vijande:2009ac}, we find
\begin{align}
    \<\f_J\|H\|\f_J\>&=\M+\frac{8}{3}\left(C_{QQ}+C_{qq}\right)+\frac{8}{3}C_{Q\q}\left[J(J+1)-4\right]\, ,\label{eq:me1}\\
    \<\f_0'\|H\|\f_0'\>&=\M+4(C_{QQ}+C_{qq})\, ,\label{eq:me2}\\
    \<\f_0'\|H\|\f_0\>&=-8\sqrt 6 C_{Q\q}\, ,\label{eq:me3}
\end{align}
which is consistent with the results of Refs.~\cite{Buccella:2006fn,Wu:2016vtq}. Note that $C_{QQ}$ and $C_{qq}$ appear only in the combination $C_{QQ}+C_{qq}$, and because of this, it turns out to be convenient to introduce the dimensionless ratio
\begin{align}
    R=\frac{2C_{Q\q}}{C_{QQ}+C_{qq}}\label{eq:r}
\end{align}
which, in the case of $QQ\Q\Q$ states, reduces to the simpler ratio of $Q\Q$ and $QQ$ couplings,
\begin{align}
    R=\frac{C_{Q\Q}}{C_{QQ}}\, .\label{eq:rr}
\end{align}
If the couplings are parameterised as in equation~\eqref{eq:cij1}, then for two-flavour states ($QQ\q\q$), $R$ depends only on the ratio of quark masses $m_q$ and $m_Q$,
\begin{align}
    R=\frac{2m_q/m_Q}{1+(m_q/m_Q)^2} \, ,
\label{eq:rratio}
\end{align}
and takes values in the range $0<R<1$, while for one-flavour states ($QQ\Q\Q$) obviously $R=1$.

As discussed, the spectrum of the diquark model comes from truncating the basis to include only the hidden colour-triplet states, namely $\f_2$, $\f_1$ and $\f_0$, but not $\f_0'$. Evaluating equation~\eqref{eq:me1}, the masses of the tensor ($M_2$), axial ($M_1$) and scalar ($M_0$) are
\begin{align}
    M_2&=\M+\frac{8}{3}\left(C_{QQ}+C_{qq}\right)(1+R)\, ,\label{eq:m2}\\
    M_1&=\M+\frac{8}{3}\left(C_{QQ}+C_{qq}\right)(1-R)\, ,\label{eq:m1}\\
    M_0&=\M+\frac{8}{3}\left(C_{QQ}+C_{qq}\right)(1-2R)\, .\label{eq:mf0}
\end{align}

To get results for the quark model, we expand the basis to include $\f_0'$ which implies, as discussed previously, that the spectrum includes two scalar states. The masses of the tensor ($M_2$) and axial ($M_1$) are as above, but the masses of the scalars ($M_0$ and $M_0'$) are the eigenvalues of
\begin{align}
    H=\M+\left(C_{QQ}+C_{qq}\right)\begin{pmatrix}\frac{8}{3}(1-2R)&-4\sqrt 6 R\\-4\sqrt 6 R&4\end{pmatrix},\label{eq:h:chromo22}
\end{align}
namely
\begin{align}
    M_0&=\M+\frac{2}{3}\left(C_{QQ}+C_{qq}\right)\left(5-4R-\Delta\right),\label{eq:m0}\\
    M_0'&=\M+\frac{2}{3}\left(C_{QQ}+C_{qq}\right)\left(5-4R+\Delta\right),\label{eq:m0p}
\end{align}
where
\begin{align}
    \Delta=\sqrt{232R^2+8R+1} \, ,
\end{align}
and we are adopting the convention that $M_0'>M_0$. The eigenstates $\psi_0$ and $\psi_0'$ corresponding to masses $M_0$ and $M_0'$ are orthogonally mixed
\begin{align}
\|\psi_0\>&=\cos\theta\|\f_0\>+\sin\theta\|\f_0'\>\, ,\\ \|\psi_0'\>&=-\sin\theta\|\f_0\>+\cos\theta\|\f_0'\>\,,
\end{align}
with mixing angle
\begin{align} \label{angle}
    \theta= \tan^{-1}\left(\frac{\Delta-1-4R}{6\sqrt 6 R}\right).
\end{align}

The mass formulae above imply unambiguous orderings for the masses of states, regardless of parameters. For diquark models, the ordering is 
\begin{align}
    M_0<M_1<M_2\,,\label{eq:ordering1}
\end{align}
whereas in quark models it is
\begin{align}
    M_0<M_1<M_2<M_0' \,. \label{eq:ordering2}
\end{align}
This can help to assign quantum numbers to experimental candidates, as discussed in Ref.~\cite{Anwar:2023a}. 

The results in this section are exact for (quark or diquark) models with effective masses, whereas for potential models (whether quark or diquark), they apply in the limit of perturbation theory. In the particular case of the quark potential model, there is an additional caveat: recalling the discussion at the end of Section~\ref{sec:quark},  the results derived above are valid only subject to the additional assumption that the spatial wavefunction of the tetraquark is totally symmetric under the interchange $Q\leftrightarrow\q$ (or $Q\leftrightarrow\Q$, in the one flavour case). As discussed, in many papers this assumption applies (even if implicitly), and we have found that in such cases (for example Refs.~\cite{Liu:2019zuc,Zhang:2022qtp}) the results agree with all of our results above (for masses, mixing angles, and mass orderings). In quark model studies which do not use that assumption, there are some differences, which are immediately apparent in violations of the mass ordering \eqref{eq:ordering2}, for example in Ref.~\cite{Wang:2019rdo}.

There is an intriguing connection between quark and diquark models in the limit of small $R$. In this limit $\Delta\approx 1+4R$, meaning the scalar masses are
\begin{align}
    M_0&\approx\M+\frac{8}{3}\left(C_{QQ}+C_{qq}\right)(1-2R)\,,\label{eq:mlim1}\\
    M_0'&\approx\M+4\left(C_{QQ}+C_{qq}\right)\,, \label{eq:mlim2}
\end{align}
which is equivalent to a perturbative treatment of the Hamiltonian~\eqref{eq:h:chromo22} to first order in $R$. Note that the lighter scalar $M_0$ reproduces the result of the diquark model, equation~\eqref{eq:mf0}. In the same limit $\theta\approx 0$ so the lighter scalar is purely $\f_0$, again coinciding with the diquark model result. So apart from the existence of a heavier scalar state, we have found that spectra of the quark and diquark models coincide in the limit of small $R$, both in  terms of masses and wavefunctions. With reference to the definitions~\eqref{eq:r} and \eqref{eq:rr}, small $R$ means that quark-antiquark interactions are small compared to quark-quark interactions. That is precisely the limit in which the diquark concept is physically reasonable.

From equation \eqref{eq:rratio}, the small $R$ limit applies to $QQ\q\q$ tetraquarks with $m_Q\gg m_q$. On this basis we suggest that the diquark model is a sensible approximation to the quark model in such cases (modulo the absent heavier scalar state). Otherwise, the spectra of the quark and diquark models are rather different, and we explore this further in the next section. The one-flavour case $QQ\Q\Q$ actually deviates maximally from the small $R$ limit, as it has $R=1$, which is the upper limit on $R$ from equation~\eqref{eq:rratio}.

Returning to the $QQ\q\q$ case, it is clear from above that as $m_Q\to\infty$, the lighter scalar decouples from $\f_0'$ and becomes purely $\f_0$. This effect has been discussed previously for $QQ\q\q$ states~\cite{Heller:1986bt,Vijande:2003ki}. In terms of colour, it is exactly what was observed also for the isoscalar $QQ\u\d$ tetraquark~\cite{Hernandez:2019eox}: as $m_Q\to\infty$, the ground state decouples from $\six\otimes\bar\six$ and becomes purely $\bar\three\otimes\three$. The comparison suggests that the effect may be generic, noting that in spite of its apparent similarity, the isoscalar $QQ\u\d$ system is very different to our $QQ\q\q$ system, because of the isospin asymmetry, which implies different spin-colour configurations for $\u\d$ compared to $\q\q$.

\section{Mass relations}
\label{sec:massrelations} 
The mass formulae in the previous section imply relations among the masses of the states, and as far as we know these have not yet been identified in the literature. For diquark models, the situation is very simple: equations \eqref{eq:m2}, \eqref{eq:m1} and \eqref{eq:mf0} imply that the masses $M_2$, $M_1$, $M_0$ satisfy the following linear relation,
\begin{align}
M_1=\frac{1}{3}\left(2M_0+M_2\right),\label{eq:diq:relation}
\end{align}
independently of model parameters. 

For quark models, the situation is only slightly more complicated. From equations \eqref{eq:m2}, \eqref{eq:m1}, \eqref{eq:m0}, and \eqref{eq:m0p}, it is clear that any mass splitting among  $M_2$, $M_1$, $M_0$ and $M_0'$ is independent of $\M$, while any ratio of such splittings is independent of $C_{QQ}+C_{qq}$, leaving a function of $R$ only. By taking ratios of different combinations of mass splittings, we get linear relations among the masses, similar to equation \eqref{eq:diq:relation} for the diquark model, but in this case involving $R$. In this way, we can find a linear relation among any combination of three masses out of the four ($M_2$, $M_1$, $M_0$ and $M_0'$), meaning a total of four relations. We concentrate on the following two:
\begin{align}
    M_1&=M_0+\frac{\Delta-1}{\Delta-1+8R}(M_2-M_0)\,,\label{eq:relation1}\\
    M_0'&=M_0+\frac{2\Delta}{\Delta-1+8R}(M_2-M_0)\,,\label{eq:relation2}
\end{align}
noting that the first of these is the closest analogue of the diquark model result~\eqref{eq:diq:relation}, in the sense of offering a formula for $M_1$ in terms of $M_0$ and $M_2$. Indeed, it reduces to the diquark model result \eqref{eq:diq:relation}, in the limit of small $R$ (taking $\Delta\approx 1+4R$). This reinforces our previous observation that the quark model reduces (apart from the heavier scalar) to the diquark model, in the limit of small $R$.

Ultimately the utility of these mass relations is that, given any two experimental candidates, we may predict the mass of the other state (in diquark models), or the other two states (in quark models). 
Since these predictions are independent of (or only weakly dependent on) parameters, they provide a very direct test of models, which can be checked against future experimental data. In Ref.~\cite{Anwar:2023a} we apply this approach to a putative multiplet of $cc\c\c$ states observed at LHCb, CMS and ATLAS. 

\begin{figure}
    \centering
    \includegraphics[width=\textwidth]{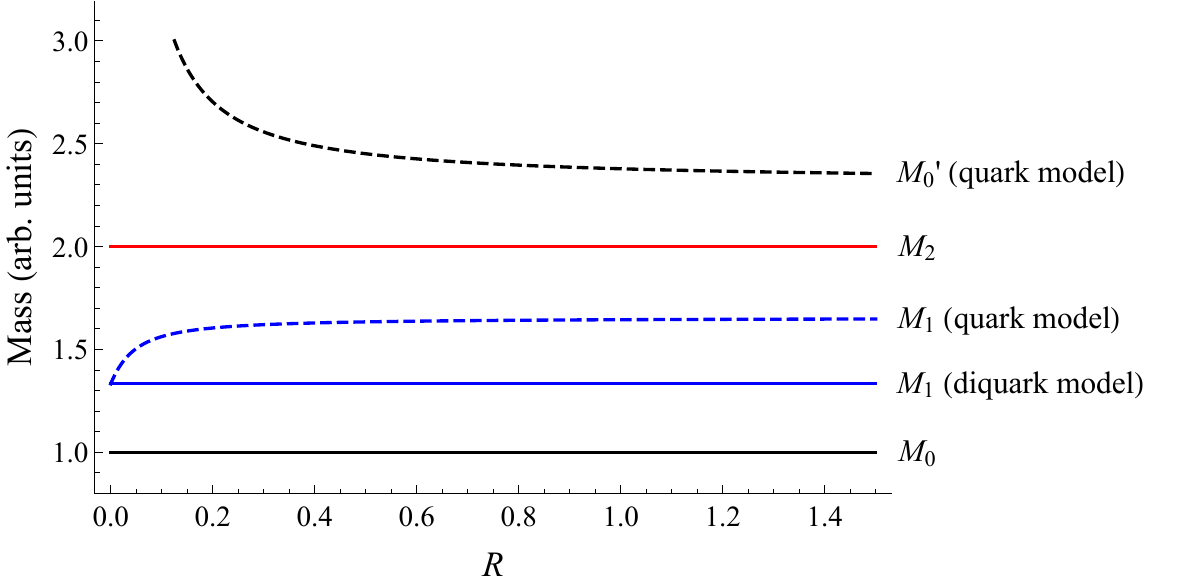}
    \caption{The mass spectrum (in arbitrary units) as a function of $R$, where $M_0$ and $M_2$ are fixed, $M_1$ (diquark model) is given by equation~\eqref{eq:diq:relation}, while
    $M_1$ (quark model) and $M_0'$ (quark model) are given by equations~\eqref{eq:relation1} and \eqref{eq:relation2}, respectively.}
    \label{fig:spectrum}
\end{figure}

The relations also lead to a simple and very general understanding of the pattern of masses which characterise quark and diquark models. In Figure~\ref{fig:spectrum} we show the mass spectrum in arbitrary units, having fixed $M_0$ and $M_2$, and using the relations \eqref{eq:diq:relation}, \eqref{eq:relation1} and \eqref{eq:relation2} to predict $M_1$ and $M_0'$. Note of course that in quark models, the pattern of masses in the multiplet is sensitive to $R$, whereas in diquark models, it is not. As a reminder, if the couplings $C_{ij}$ are parameterised as in equation~\eqref{eq:cij1}, then $QQ\Q\Q$ states have $R=1$, while $QQ\q\q$ states have $0<R<1$, with $R\approx 0$ for $m_Q\gg m_q$. Notice in Figure~\ref{fig:spectrum} that as $R\to 0$, the scalar masses in the quark and diquark models become degenerate, as anticipated above.

Comparison of the spectrum in Fig.~\ref{fig:spectrum} to experimental candidates can help to reveal the underlying dynamics of a multiplet of tetraquark states. In particular, apart from in the limit of small $R$, quark and diquark models can be differentiated by the relative position of the axial ($M_1$) compared to the scalar ($M_0$) and tensor ($M_2$). The difference is particularly pronounced for larger $R$, including $R=1$ which applies to the single flavour case ($QQ\Q\Q$). 

We have been assuming that the couplings satisfy equation~\eqref{eq:cij1}, but empirically this is not universally reliable; see for example the couplings fitted to mesons and baryons in Refs.~\cite{Buccella:2006fn,Cheng:2020wxa}. Moreover, in quark potential models, it applies only in a symmetry limit which is not strictly justified by the Hamiltonian (see Section~\ref{sec:quark}). 
If we no longer assume equation~\eqref{eq:cij1}, 
then of course $R$ is no longer constrained to $0<R\le 1$, and for this reason in Figure~\ref{fig:spectrum} we extend the plot to larger values of $R$. 

Without equation~\eqref{eq:cij1}, for one-flavour states ($QQ\Q\Q$) it is no longer true that $R=1$ exactly. Using the more general definition of $R$ in equation~\eqref{eq:rr}, we expect deviations from $R=1$ due to asymmetry in $QQ$ and $Q\Q$ spatial wavefunctions. In this context it is reassuring that the spectrum in Fig.~\ref{fig:spectrum} is not very sensitive to the choice of $R$, for values near $R=1$.

\begin{figure}
\centering
\begin{tikzpicture}
\node (image) 
{
  \includegraphics[width=0.68\textwidth,height=0.62\textwidth]{{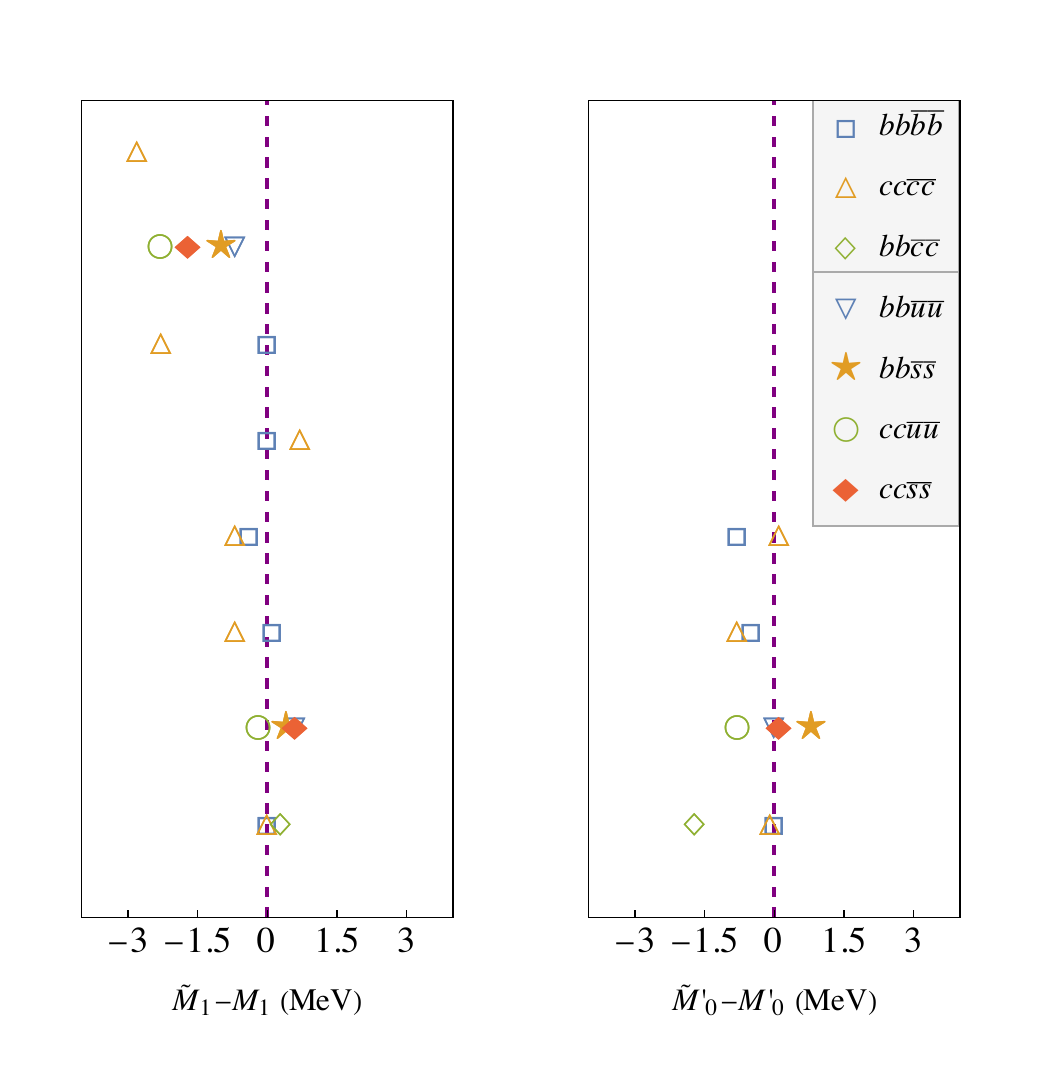}}
};
\node[xshift=2.5cm,yshift=0cm]
[
    align=center,
    text=black,
    font={\scriptsize}
] at (-2.5, 3.7)
{DPM\,\cite{Debastiani:2017msn}};
\node[xshift=2.5cm,yshift=0cm]
[
    align=center,
    text=black,
    font={\scriptsize}
] at (-2.5, 2.7)
{DPM\,\cite{Ebert:2007rn}};
\node[xshift=2.5cm,yshift=0cm]
[
    align=center,
    text=black,
    font={\scriptsize}
] at (-2.5, 1.8)
{DPM\,\cite{Lundhammar:2020xvw}};
\node[xshift=2.5cm,yshift=0cm]
[
    align=center,
    text=black,
    font={\scriptsize}
] at (-2.5, 0.9)
{DM\,\cite{Berezhnoy:2011xn}};
\node[xshift=2.5cm,yshift=0cm]
[
    align=center,
    text=black,
    font={\scriptsize}
] at (-2.5, -0.05)
{QPM\,\cite{Liu:2019zuc}};
\node[xshift=2.5cm,yshift=0cm]
[
    align=center,
    text=black,
    font={\scriptsize}
] at (-2.5, -1)
{QPM\,\cite{Zhang:2022qtp}};
\node[xshift=2.5cm,yshift=0cm]
[
    align=center,
    text=black,
    font={\scriptsize}
] at (-2.5, -1.85)
{CQM\,\cite{Cheng:2020wxa}};
\node[xshift=2.5cm,yshift=0cm]
[
    align=center,
    text=black,
    font={\scriptsize}
] at (-2.5, -2.8)
{CQM\,\cite{Wu:2016vtq}};
\end{tikzpicture}
\caption{The left and right panels show, for the axial and heavy scalar, respectively, the difference between the mass obtained from the relations \eqref{eq:diq:relation}, \eqref{eq:relation1} and \eqref{eq:relation2}, and the quoted mass taken from the literature, including examples of all four classes of model: the chromomagnetic quark model~(CQM), quark potential model (QPM), diquark model (DM), and diquark potential model~(DPM). }
\label{fig:relationcheck}
\end{figure}
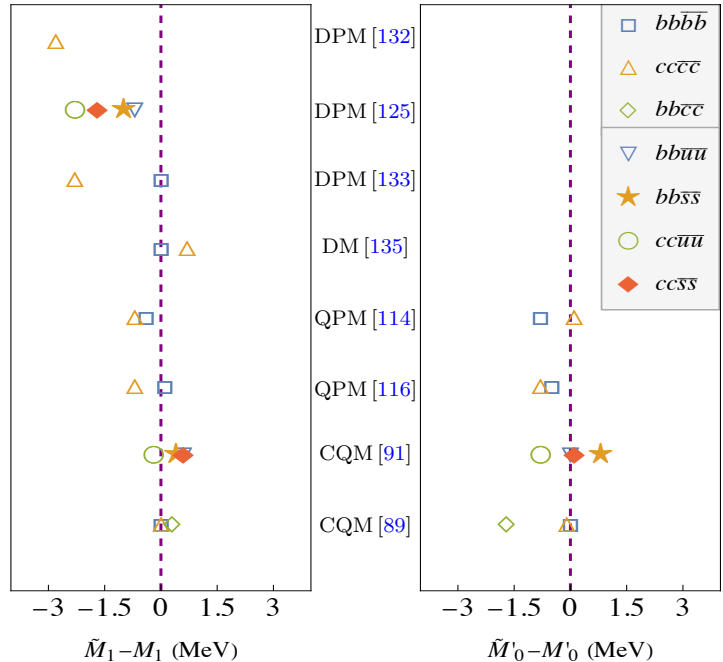

Although the relations \eqref{eq:diq:relation}, \eqref{eq:relation1} and \eqref{eq:relation2} have seemingly not been discussed previously, they are actually apparent in the quoted mass predictions throughout the literature, for each of the four classes of model we consider in this paper. In the Table~\ref{tab:masscheck} (Appendix) we compile the masses $M_0$, $M_1$, $M_2$ and $M_0'$ quoted in many different model calculations. To check the validity of the relations, for each model we have taken  $M_0$ and $M_2$ as inputs and, using the mass relations, we have computed the masses of the axial $\widetilde M_1$ and (if appropriate) the heavy scalar $\widetilde M_0'$, which we can then compare to the corresponding quoted values of $M_1$ and $M_0'$. In the last two columns in Table~\ref{tab:masscheck}, and also in Figure~\ref{fig:relationcheck}, we show the differences $\widetilde M_1-M_1$ and $\widetilde M_0'-M_0'$, which in most cases are 1~MeV or less. This is a striking confirmation of the validity of the relations, across all classes of model.

From previous discussions, we know that the mass relations are exact for (quark and diquark) models based on effective masses (corresponding to Refs.~\cite{Wu:2016vtq, Cheng:2020wxa,Berezhnoy:2011xn,Berezhnoy:2011xy} in Table~\ref{tab:masscheck} and Figure~\ref{fig:relationcheck}). In such cases, where $\widetilde M_1-M_1$ or $\widetilde M_0'-M_0'$ deviate from zero, this can be attributed to rounding errors, from having applied the relations using inputs which are quoted to a particular number of significant figures. 

In (quark and diquark) potential models, the relations apply strictly in the limit of perturbation theory. Many of the masses in Table~\ref{tab:masscheck} and Figure~\ref{fig:relationcheck} have been computed in perturbation theory, so we would expect the mass relations to be satisfied exactly, up to small rounding errors as mentioned previously. Notably, the masses of Ref.~\cite{Debastiani:2017msn} are not computed in perturbation theory, and the somewhat larger deviation from exact agreement in this case can be understood for that reason.

In the specific case of quark (not diquark) potential models, we recall the additional caveat mentioned previously: our mass formulae -- hence the resulting mass relations -- apply only subject to the assumption that the wavefunction is symmetric under the interchange $Q\leftrightarrow \q$ (or $Q\leftrightarrow \Q$, in the one flavour case). The particular examples shown  in Table~\ref{tab:masscheck} and Figure~\ref{fig:relationcheck} (Refs. \cite{Liu:2019zuc,Zhang:2022qtp}) satisfy this requirement, because in their implementation of the Gaussian Expansion Method, the symmetry is automatic for degenerate quarks ($cc\c\c$ and $bb\b\b$). This is to some extent an artefact of the calculation, since the symmetry is not actually imposed by the symmetries of the Hamiltonian (see Section \ref{sec:quark}). In models where the symmetry is not imposed (such as Refs.~\cite{Wang:2019rdo,Lu:2020cns}) the relations are not satisfied. Effectively this is because the chromoelectric term in the Hamiltonian induces a splitting between the $\f_{0,1,2}$ and $\f_0'$ states which, before spin splitting, would otherwise be degenerate. For precisely the same reason, the relations also do not apply to the extended chromomagnetic model of Refs.~\cite{Weng:2020jao,Guo:2021yws} which, unlike the ordinary chromomagnetic model, have a chromoelectric splitting in the center of mass term.

\section{Conclusions}
\label{sec:conclusion}

One of the main obstacles to progress in understanding the nature of exotic hadrons is that models are not very well constrained.  This is because of the intrinsic ambiguity in identifying the relevant degrees of freedom and their interactions, and also because model parameters are only weakly constrained by comparison with the spectra of conventional hadrons. Consequently, absolute mass predictions for tetraquark states are subject to systematic uncertainties which are large and difficult to quantify, so comparing these to experimental candidates can hardly discriminate among models.

Our perspective is that it is considerably more useful to examine not absolute mass predictions, but relations among masses. As well as being more reliable -- in the sense of depending only weakly on model parameters, or not at all -- such relations are considerably more direct, and therefore effective, as a way of discriminating among competing models.

The most important distinguishing feature of models is whether they include all colour configurations (quark models) or only a subset (diquark models). For each class of model, we showed that the corresponding potential model is equivalent (in perturbation theory) to a simpler model with effective (quark or diquark) masses -- though in the case of quark models, the equivalence relies on an assumption of spatial symmetry which, though commonplace, is not strictly justified. 

We derived general formulae for the mass spectrum of S-wave $QQ\q\q$ and $QQ\Q\Q$ states in quark and diquark models, and showed how the two models coincide in an appropriate limit. From the formulae, we identified several resulting linear relations which are independent of,  or only weakly dependent on, model parameters. The relations are exact for (quark or diquark) models  with effective masses, or valid in perturbation theory for (quark or diquark) potential models. Although the relations have seemingly not been discussed in the literature before, they are apparent in the quoted mass predictions in all classes of models. 

The relations reveal how quark and diquark models have a  characteristically different pattern of masses, which can be tested against future experimental data. In particular, given any two experimental candidates, using the relations one can predict the masses of the additional one or two states (in diquark or quark models, respectively). In a forthcoming paper \cite{Anwar:2023fbp} we apply this concept (and some other results from the present work) to the apparent $cc\c\c$ states observed at LHCb, CMS and ATLAS.

\begin{acknowledgments}
This work is supported by The Royal Society through Newton International Fellowship.
\end{acknowledgments}


\newpage

\appendix
\section*{Appendix}

\begin{table}[h!]
\def\arraystretch{1.17}
\centering
    \begin{tabularx}{\textwidth}{lXrrrrcc}
    \hline
			&		&	$M_0~~$	&	$M_1~~$	&	$M_2~~$	&	$M_0'~~$	&	$\widetilde M_1-M_1$	&	$\widetilde M_0'-M_0'$	\\
\hline														Chromomagnetic quark model	\cite{Wu:2016vtq}	&	$bb\b\b$	&	20155.4	&	20211.6	&	20242.5	&	20275.5	&$	0.0	$&$	0.0	$\\
		&	$cc\c\c$	&	6796.6	&	6899.2	&	6955.7	&	7016	&$	0.0	$&$	-0.1	$\\
		&	$bb\c\c$	&	13496.5	&	13559.7	&	13594.9	&	13633.8	&$	0.3	$&$	-1.7	$\\
\hline																								
Chromomagnetic quark model	\cite{Cheng:2020wxa}	&	$bb\u\u$	&	10642	&	10676	&	10699	&	10738	&$	0.6	$&$	0.0	$\\
		&	$bb\bar s\bar s$	&	10858	&	10901	&	10926	&	10954	&$	0.4	$&$	0.8	$\\
		&	$cc\u\u$	&	4000	&	4124	&	4194	&	4277	&$	-0.2	$&$	-0.8	$\\
		&	$cc\bar s\bar s$	&	4227	&	4358	&	4430	&	4502	&$	0.6	$&$	0.1	$\\
\hline																
Quark potential model	\cite{Zhang:2022qtp}	&	$bb\b\b$	&	19200	&	19216	&	19225	&	19235	&$	0.1	$&$	-0.5	$\\
		&	$cc\c\c$	&	6411	&	6453	&	6475	&	6500	&$	-0.7	$&$	-0.8	$\\
\hline																
Quark potential model	\cite{Liu:2019zuc}	&	$bb\b\b$	&	19306	&	19329	&	19341	&	19355	&$	-0.4	$&$	-0.8	$\\
		&	$cc\c\c$	&	6455	&	6500	&	6524	&	6550	&$	-0.5	$&$	0.1	$\\
\hline																
Diquark model	\cite{Berezhnoy:2011xn,Berezhnoy:2011xy}	&	$bb\b\b$	&	18754	&	18808	&	18916	&		&$	0.0	$&$		$\\
		&	$cc\c\c$	&	5966	&	6051	&	6223	&		&$	0.7	$&$		$\\
\hline																
Diquark potential model	\cite{Lundhammar:2020xvw}	&	$bb\b\b$	&	18723	&	18738	&	18768	&		&$	0.0	$&$		$\\
		&	$cc\c\c$	&	5960	&	6009	&	6100	&		&$	-2.3	$&$		$\\
\hline																
Diquark potential model	\cite{Ebert:2007rn}	&	$bb\u\u$	&	10648	&	10657	&	10673	&		&$	-0.7	$&$		$\\
		&	$bb\bar s\bar s$	&	10932	&	10939	&	10950	&		&$	-1.0	$&$		$\\
		&	$cc\u\u$	&	4056	&	4079	&	4118	&		&$	-2.3	$&$		$\\
		&	$cc\bar s\bar s$	&	4359	&	4375	&	4402	&		&$	-1.7	$&$		$\\
\hline																
Diquark potential model	\cite{Debastiani:2017msn}	&	$cc\c\c$	&	5969.4	&	6020.9	&	6115.4	&		&$	-2.8	$&$		$\\
\hline

\end{tabularx}
    \caption{The masses $M_0$, $M_1$, $M_2$ and $M_0'$ are taken from literature calculations in different models, whereas $\widetilde M_1$ and $\widetilde M_0'$ are computed from the mass relations \eqref{eq:diq:relation}, \eqref{eq:relation1} and \eqref{eq:relation2}, having taken $M_0$ and $M_2$ as inputs. In the last two columns we show the differences $\widetilde M_1-M_1$ and $\widetilde M_0'-M_0'$, as a measure of the accuracy of the mass relations. }
    \label{tab:masscheck}
    \end{table}

\newpage


%

\end{document}